# Production, properties and potential of graphene

Caterina Soldano, Ather Mahmood[§] and Erik Dujardin[1]

CEMES, Centre d'Elaboration de Matériaux et d'Etudes Structurales - CNRS UPR 8011

29 rue Jeanne Marvig – B.P. 94347 - 31055 Toulouse, Cedex 4

Tel: (+33) (0)5 62 25 78 38. Fax: (+33) (0)5 62 25 79 99

[§] Present address: Institut Néel, CNRS-UJF, BP 166, 38042 Grenoble Cedex 9, France.

Outline

[1] Corresponding author. Fax: +33 562257999.  E-mail address: erik.dujardin@cemes.fr (E. Dujardin)

Abstract: This review on graphene, a one atom thick, two-dimensional sheet of carbon atoms, starts with a general description of the graphene electronic structure as well as a basic experimental toolkit for identifying and handling this material. Owing to the versatility of graphene properties and projected applications, several production techniques are summarized, ranging from the mechanical exfoliation of high quality graphene to the direct growth on carbides or metal substrates and from the chemical routes using graphene oxide to the newly developed approach at the molecular level. The most promising and appealing properties of graphene are summarized from an exponentially growing literature, with a particular attention to matching production methods to characteristics and to applications. In particular, we report on the high carrier mobility value in suspended and annealed samples for electronic devices, on the thickness-dependent optical transparency and, in the mechanical section, on the high robustness and full integration of graphene in sensing device applications. Finally, we emphasize on the high potential of graphene not only as a post-silicon materials for CMOS device application but more ambitiously as a platform for post-CMOS molecular architecture in electronic information processing.

# 1. Introduction

When Lavoisier listed "*Carbone*" in his "*Traité Elémentaire de Chimie*" as one of the newly identified chemical elements, for the first time just 220 years ago, he had already identified the versatility of carbon since he had shown that it was the elementary component of both diamond and graphite.[1][2] Since then, more allotropes of carbon have been reported and a large scientific community has been passionate about deciphering the properties of this element that can adopt many structures ranging from diamond and graphite (3D), graphene (2D),[2] nanotubes (1D) [3] or fullerenes (0D) [4] as illustrated in Figure 1. The former three-dimensional allotropes have been known and widely used for centuries, whereas fullerenes and nanotubes have been only discovered and studied in the last two decades. With the exception of diamond, it is possible to think of fullerenes, nanotubes and graphite as different structures built from the same hexagonal array of $sp^2$ carbon atoms, namely graphene. Indeed, fullerenes and nanotubes can be mentally visualized as a graphene sheet rolled into a spherical and cylindrical shape, respectively, and graphite can be described as a stack of alternately shifted graphene sheets.

For many years graphene was a mere concept used for this descriptive approach of more complex forms of aromatic carbon and an appealingly simple system for solid state theoreticians. Except for the early observations of chemically derived graphene sheets (see Section 3.3) and, later on, the characterization of graphite monolayers on metals and carbide (see Section 3.2), very few experimentalists considered graphene as of interest *per se* until 2004.[5] However, the direct observation of isolated graphene monolayer on that year has sparked an exponentially growing interest and just a few years were enough to gather several scientific communities to investigate the properties of this new yet ancient two-dimensional material. Graphene has first attracted the curiosity of mesoscopic physicists owing to its peculiar electronic behavior under magnetic field and at low temperature. The investigation and tailoring of its transport properties from macroscopic to molecular scales captures a large share of the current research effort. Technologists and materials scientists have rapidly grabbed some of the assets of graphene and are already exploring the ways of incorporating graphene into applied devices and materials. Eventually, chemists and surface physicists remembered their ancient recipes which could provide solutions to mass produce the elusive carbon monolayer either in bulk suspensions or in substrate-supported forms.

Considering the recent move of a large variety of communities converging to graphene-related research topic, we have opted for a panoramic approach to this Review article rather than a focused opus, which would have been another viable option if one considers the more than 3000 articles published in this 5-year span. Where necessary, the Reader will be redirected to more specific recent reviews. After some simple background information on graphene, the basic set of techniques needed to identify graphene and probe its established properties are described in Section 2. Section 3 describes how to produce graphene by physical and chemical approaches, each of which being commented in terms of graphene quality and morphology as well as the specific uses for which it is suitable. Section 4 is dedicated to the description of recently discovered or characterized electronic, optical and mechanical properties, which allows to extract some perspectives in Section 5. Far from being exhaustive, but through a series of examples, this Review article would like to convey some of the uplifting enthusiasm that has gained the now well-identified but still growing graphene research community.

---

[2] Text available at http://www.lavoisier.cnrs.fr/

## 2. Overview and characterization techniques

Graphene is composed of $sp^2$-bonded carbon atoms arranged in a two-dimensional honeycomb lattice as shown in the Figure 2(a). The lattice can be seen as consisting of two interpenetrated triangular sub-lattices, for which the atoms of one sub-lattice are at the centers of the triangles defined by the other with a carbon-to-carbon inter-atomic length, $a_{C-C}$, of 1.42 Å. The unit cell comprises two carbon atoms and is invariant under a rotation of 120° around any atom. Each atom has one $s$ orbital and two in-plane $p$ orbitals contributing to the mechanical stability of the carbon sheet. The remaining $p$ orbital, perpendicularly oriented to the molecular plane, hybridizes to form the $\pi^*$ (conduction) and $\pi$ (valence) bands, which dominate the planar conduction phenomena.[6] The energies of those bands depend on the momentum of the charge carriers within the Brillouin zone. In the low energy regime, i.e. in the vicinity of the K and K' points, those two bands meet each other producing conical-shaped valleys, as shown in Figure 2(b). In this low-energy limit, the energy-momentum dispersion relation is linear (relativistic Dirac's equation) and carriers are seen as zero-rest mass relativistic particles with an effective "speed of light", $c^* \sim 10^6$ m/s. The peculiar local electronic structure, thus allows the observations of numerous phenomena which are typical of a two-dimensional Dirac fermions gas (See Section 4.1).[7] In the high-energy limit, the linear energy-momentum relation is no longer valid and the bands are subjected to a distortion leading to anisotropy, also known as trigonal warping.[6] Upon stacking layers on top of each other, one first obtains bilayered graphene, which exhibits its own set of very specific properties. The center of the aromatic rings of the upper graphene sheet sits on top of an atom of the lower sheet, so that the symmetry is trigonal rather than hexagonal. With the inter-plane interaction, the charge carriers acquire a mass and the dispersion recovers a parabolic dispersion described by the Schrödinger formalism. Nevertheless, bilayer remains gapless if one ignores trigonal warping. The interaction of the two $\pi$ and $\pi^*$ bands of each graphene sheets produces two other bands. In the presence of an external potential, a gap can be opened close to the K point.[8, 9] Further generalization of the perfect stacking results in graphite with alternating (ABAB, Bernal type) or staggered (ABCABC, rhombohedral type) arrangement depending on whether the lateral carbon atom shift changes direction from one layer to the next or not. The interplanar distance of ideal graphite is 3.45 Å but if successive planes are rotated with respect of each other, this spacing can increase. For the electronic structure of graphite, one can refer to [10].

While graphene has been known as a textbook structure to calculate band diagrams and predict unique electronic properties since the early 1940s, the experimental investigation of graphene properties, as a standalone object, has been almost inexistent until the very recent years because of the difficulty to identify and univocally characterize the single-atom thick sheet. Before detailing the graphene production methods in Section 3 and its physical properties in Section 4, it is appropriate to provide the reader with the tools to locate, recognize and characterize graphene. Rather than a complete list and description of relevant techniques, a basic toolkit is summarized here, that any team working on graphene should get access to in order to facilitate graphene-based projects. Therefore, Figure 3 presents graphene identification data from (a) optical, (b) scanning probe and (c) electron microscopy as well as (d) ARPES, (e) Raman and (f) Rayleigh spectroscopy techniques.

**2.1 Optical microscopy.** One of the key elements in graphene discovery, and still nowadays in handling it in most studies, resides in the use of an appropriate substrate that maximizes the optical contrast of the carbon atom monolayer in the wavelength range of maximal sensitivity for the experimentalist. In the case of commonly used oxide-covered silicon wafers, the deposited graphene adds a small optical path to the Fabry-Perrot cavity created by the $SiO_2$ layer on silicon.[11] By adjusting the silica thickness to 90 or 300 nm, the reflected light intensity is maximal at about 550 nm, i.e. at the maximum of the human eye sensitivity.[2] The short optical path added by graphene as thin as a monolayer can be easily seen since the contrast between graphene and the substrate can then be as high as 12%. The identification of graphene samples having lateral sizes of a couple of microns is thus greatly facilitated (Figures 3a and 4). Alternatives to silica have been studied and graphene is also visible on 50 nm $Si_3N_4$ using blue light [11], 72 nm $Al_2O_3$ on Si wafer [12], and on 90 nm polymethyl-methacrylate (PMMA) using white light. [11, 13] Furthermore, the optical contrast of graphene or few-graphene layers can be noticeably improved by suitably adjusting the monochromatic light wavelength at the maximal contrast.[11, 14] The optimization of the optical contrast of graphene with respect to the background colour of the substrate has also been achieved by a number of image processing methods. The simplest method, which is only suitable for 300-nm thick oxide where the maximal contrast occurs at a wavelength of 550 nm, consists in displaying the green channel image normalized by its value on the bare substrate. A more sophisticated approach, which takes into account the whole spectrum of the white light, has been fully explained in reference [12]. However, so far, these numerical approaches remain offline image processing which do accelerate the identification process of graphene but still lack real-time user friendliness.

**2.2 Atomic Force Microscopy (AFM).** AFM was quite naturally one of the first techniques used to establish that the thinnest graphitic flakes first identified optically were indeed one-atom thick monolayers. However, a question immediately arises: what should be the apparent height of graphene on a $SiO_2$ substrate with a typical 1-nm r.m.s. roughness? A single layer of graphene on crystalline graphite shows a typical 0.4 nm thickness in intermittent contact AFM mode. Surprisingly, a single layer graphene on oxidized wafers consistently appears to be 0.8 to 1.2 nm thick with any supplementary layer on top of it adding the expected 0.35 nm thickness, which corresponds to the native van der Waals interlayer distance.[15, 16] The uncertainty about the nature of 0.8-nm thick graphene objects was lifted by studying self-folded sheets,[17] as shown in Figure 3b, or by combining AFM with micro-Raman spectroscopic data.[18] The origin for the extra apparent thickness remains unclear since pure van der Waals interaction between silica and graphene can not account for it. Part of the answer could lie in the protocol used to deposit graphene in ambient conditions. When a piece of single layered graphene is manipulated by an AFM tip, it reveals a footprint that has exactly the same shape and an apparent height of 0.3 to 0.6 nm depending on the AFM ambient humidity. Such an imprint is probably composed of atmospheric hydrocarbons between the graphene and silica surfaces by capillary condensation.[3] Although AFM is too slow and limited in lateral scan size to be used as a primary identification method and in spite of the extra apparent height, AFM soft imaging modes are the best way to monitor the topological quality of substrate-supported graphene samples during the successive steps of device processing.

**2.3 Transmission Electron Microscopy (TEM).** With the advent of colloidal methods to produce graphene sheets as well as the improvement of device characteristics in unsupported graphene, TEM has recently arisen as a structural characterization tool for suspended graphene that can span low magnification imaging as well as atomic scale details, as shown in Figure 3c. Moreover, single-atom thick graphene makes suspended graphene the ideal support film for high-resolution, spherical aberration-corrected TEM studies of individual defect sites [19-22] and adsorbed atoms as light as carbon and hydrogen.[23] Electron diffraction studies can be used to qualitatively distinguish a single from a bi-layer since both exhibit a six-fold symmetry, but the ratio of the intensities for the [2110] and [1100] spots is inverted [21, 24] for the two objects as predicted by Horiuchi *et al.* .[25]

**2.4 Angle-Resolved Photoemission Spectroscopy (ARPES).** Although ARPES is not a standard laboratory-scale technique, it provides a direct evidence of the electronic structure in graphene and other carbon-based materials and it should be mentioned here. When shining a substrate with 10-300 eV photons, photoelectrons are extracted from the substrate surface, the momentum and energy of which can be analyzed with as little as 15 meV resolution to reconstruct the energy band diagram (Figure 3d).[26] In the case of graphene, the relativistic Dirac-like linear dispersion near the Brillouin zone K corner [27, 28] and the chirality of the charge carriers could be directly observed as well as the emergence of small band gaps due to graphene interlayer or substrate-graphene interactions. [28-33]

**2.5 Raman scattering.** Raman scattering is a fast and non destructive technique that provides a direct insight on the electron-phonon interactions, which implies a high sensitivity to electronic and crystallographic structures. It has therefore been extensively applied to the structural investigation of carbon materials and more particularly of nanotubes.[34] Raman spectra of carbon materials show similar features in the 800-2000 cm$^{-1}$ region, which is also of interest for graphene.[35-37] The G peak at around 1560 cm$^{-1}$ corresponds to the $E_{2g}$ phonon at the centre of the Brillouin zone. The D peak, at 1360 cm$^{-1}$, is due to the out-of-plane breathing mode of the *sp*$^2$ atoms and is active in the presence of a defect.[38, 39] The D band is an efficient probe to assess the level of defects and impurities in graphene. It is for example completely silent for high quality graphene such as micromechanically exfoliated, except in proximity of the edges. However, a major fingerprint of graphene is the D' peak (sometimes labeled as 2D peak) at about 2700 cm$^{-1}$ (Figure 3e). The shape, position and intensity relative to the D band of this peak depend markedly on the number of layers.[40, 41] The isolated graphene monolayer exhibits a symmetrical Lorentzian peak centered on 2640 cm$^{-1}$ which can shift to 2655-2665 cm$^{-1}$ for loosely stacked graphene layers, as in SiC-templated graphenes.[40] For two or more layers, a second peak centered on 1685 cm$^{-1}$ emerges and eventually dominates the first peak for more than three monolayers. Beyond 5 monolayers, the Raman multi-peak profile becomes hardly distinguishable from that of bulk graphite that is a complex group of bands between 2600 and 2750 cm$^{-1}$.[35]

---

[3] Mahmood A, Linas S, Dujardin E., unpublished results.

**2.6 Rayleigh scattering.** In contrast to Raman spectroscopy, Rayleigh scattering results from the elastically scattered photons which are much more abundant. Therefore, Rayleigh scattering measurement is five orders of magnitude quicker that Raman. Rayleigh scattering has been shown to offer another quick and non-invasive method to image graphene and even identify the number of layers of a given sample (N < 6) since the monochromatic contrast varies linearly with thickness as shown in Figure 3f.[42] Contrast of 0.08 at 633 nm (He-Ne laser) with a dependence on the frequency of the incident light with a spatial resolution of 800 nm has been experimentally found.

**3. Graphene production**

Graphene has been a much debated material before it was eventually observed lying alone on a substrate. Efforts to thin graphite down to its ultimate constituent goes back to 1960, when electron microscopist Fernandez-Moran was looking for a robust, electron-beam transparent and uniform support membrane and did extract millimeter-sized graphene sheets as thin as 5 nm (~15 layers) from graphite crystals by micromechanical exfoliation.[43] Although his result became a standard method for electron microscopy sample preparation, the condensed matter physicists who had already predicted a number of properties for graphite [8] remained unaware that the single-atom thick sheet had almost been isolated. In 1962, single layers and bilayers of colloidal graphite oxide and its partially reduced form described in Section 3.3 were observed with electron microscopy by Boehm et al. for the first time.[44] The next decade witnessed a booming activity on chemical intercalation and exfoliation of oxidized graphite and other layered materials such as molybdenum or tungsten sulfides.[45, 46] Once again pure graphene monolayer would elude the sagacity of experimentalists whereas MoS$_2$ and WS$_2$ charged monolayers were identified and shown to be stable in suspension by X-ray diffraction.[47] With the discovery of fullerenes and nanotubes in the early 1990s, a renewal of interest in all kinds of carbon materials prompted not only electron but also scanning probe microscopists to consider studying graphene experimentally. AFM manipulation of freshly cleaved pyrolytic graphite yielded nanoscale origami-like structures of one graphene layer thickness but which could not be transferred from HOPG to another substrate.[5] Thin graphitic microdiscs having less than 80 monolayers were reported in 1997 [48, 49] and shortly afterwards, sub-10 nm (i.e. less than 30 monolayers) stacks of graphite were obtained by rubbing microfabricated graphite pillars on a substrate.[50] Ruoff *et al.* then suggested the possibility of producing single layer by this technique.[51] In 2004, this method was refined by Kim *et al.* who mounted a single graphite block onto an AFM cantilever to control the pressure and shearing force applied in an attempt to reach the single layer deposition conditions.[52] However, production of single layer graphene by this approach has not been reported so far.

This long and tenacious series of unsuccessful attempts to produce single sheet graphene emphasizes the timeliness as well as the ingenuity of the simple method presented by Geim's group in 2004.[2, 15] By repeatedly cleaving a graphite crystal with an adhesive tape to its limit and then by transferring the thinned down graphite onto an oxidized silicon wafer with the appropriate color, the authors have made the 2D dream come visible and hence, for all practical purposes, true and usable (Figure 4). This discovery marked the onset of experimental physics on graphene, which then made it relevant to revisit successfully all other methods to produce graphene which had reputedly failed in the past forty years. In a matter of months, the epitaxial growth of graphene layers on metal carbides (SiC in particular) by sublimation,[53, 54] or directly on metal surfaces by chemical vapor deposition (CVD) were shown to produce graphene, contrary to what had been concluded before. [55, 56] In the case of metal substrates, new methods to transfer graphene from the growth substrate to an insulating one have rendered these approaches technologically relevant.[57, 58]

In this renewed context, large scale production of graphene became a necessity and chemical routes needed to be revisited,[59] with a new challenge: reverting $sp^3$ graphene oxide to graphene by as extensive a reduction as possible. The wet approaches could be enriched by the acquired knowledge on carbon nanotube dispersion in solvents resulting in non-oxidizing solvatation routes.

Finally, the long and patient approach of synthetic chemists who are producing large polyaromatic hydrocarbons (PAH) that can be seen as sub-nanometer fragments of graphene are now taking a completely different flavor. The atomically precise structure of these small graphene molecules possesses what other methods can not offer: an absolute precision of the edge structure of nanoscale graphene domains. The challenge remains the same as it has been for 20 years: how to reach molecule size compatible with interconnecting technologies that could probe their properties?

The following sections will provide the Reader with an up-to-date overview of these four avenues towards graphene samples, and each method will be commented with respect of its assets and shortcomings, as shown in Table 1.

## 3.1 Mechanical exfoliation

With an interlayer van der Waals interaction energy of about 2 eV/nm$^2$, the order of magnitude of the force required to exfoliate graphite is about 300 nN/μm$^2$.[52] This extremely weak force can be easily achieved with an adhesive tape as experienced each time one refreshes a graphite crystal substrate for AFM or STM imaging, for example. This was first shown by rubbing microfabricated arrays of graphite micropillars.[50] One single such pillar can be micro-manipulated with a sharp glass tip to deposit it on a desired location. In a second step, the glass tip can be used to repeatedly exfoliate flakes from a graphite crystal *in situ*.[60] In a more sophisticated version, this *top-down* approach can also be performed locally by mounting a single micropillar on an AFM cantilever with purposely-chosen spring constant to control the necessary pressure and shearing force by performing force curve experiments.[52] Although these approaches provide a means to arbitrarily choose the location of the graphene deposition, they do not allow thinning the graphite crystal below a few nanometers in thickness and monolayer graphene can not be obtained.

The remarkably simple yet efficient method developed by Novoselov and Geim consists in using common adhesive tape to repeat the stick and peel process a dozen times which statistically brings a 1 μm thick graphite flake to a monolayer-thin sample. After checking for smooth and thin fragments on the tape with optical microscopy, the graphene and graphite pieces are then transferred onto the cleaned substrate by a gentle press of the tape.[2, 15] The crucial experimental aspect lies in the choice of the substrate. The apparent contrast of graphene monolayer on Si/SiO$_2$ substrate (oxide layer of either 300 or 90 nm) is indeed maximized at about 12% at 550 nm, where the sensitivity of the human eye is optimal (Figure 4). This phenomenon is easily understood by considering a Fabry-Perot multilayer cavity in which the optical path added by graphene to the interference of the SiO$_2$/Si system is maximized for specific oxide thicknesses.[11, 12, 14] Hence, in practice, thicker graphite flakes deposited on a 300 nm SiO$_2$ will appear yellow to bluish as the thickness decreases (Figure 4a), but when reaching sub-10 nm thicknesses, darker to lighter shades of purple will indicate graphene of a few and eventually one single monolayer as illustrated in Figure 4b. Since the visibility of graphene derives from an interference phenomenon in the substrate, the latter one can be changed at will by adapting the dielectric layer thickness as long as some reflectivity is preserved. One drawback of the tape technique is that it can leave glue residues on the sample, which have since then been shown to limit the carrier mobility.[61, 62] A post-deposition heat treatment, possibly in reducing atmosphere is thus required to remove the residue. Baking in hydrogen/argon environment (1h at 200°C) [63] or alternatively Joule heating under vacuum up to approximately 500°C [64] represent two routinely used methods.

To partially circumvent the glue contamination, a few attempts to promote graphene adhesion on insulating substrates by applying high voltages have been reported.[65, 66] Typically, freshly cleaved graphite is brought into contact with the deposition substrate and sandwiched together between two electrodes. A 1 to a few tens of kilovolts are applied for a few seconds and the graphite is removed from the substrate where graphene and few layer graphene samples can be found. On pure silica (300 nm on silicon wafer), applied voltages between 3 and 5 kV yielded mostly one to 3-layer thick graphene but a large thickness dispersion was observed and general cleanliness was not ensured. Interestingly, the presence of sodium ions in borosilicate improves substantially the lateral size of the deposits.[66] Based on anodic bonding, this method relies on the migration of sodium ions from the bulk to the surface after applying 1.2 to 1.7 kV while heating the substrate at 200°C. The mother crystal is then cleaved off and thinning of the sample by tape peeling yields monolayers exceeding several hundreds of microns in lateral sizes. The Raman and electrical characterization point to single graphene deposition although it is unclear if, and to which extent the anodic bonding modifies the graphene structure.

Micromechanical exfoliation remains the best method in terms of electrical and structural quality of the obtained graphene, primarily because it benefits from the high quality of the starting single crystalline graphite source. The size of the deposit is also appreciable, and can be purchased on supporting substrate in the fraction of square millimeter. However it will be challenging to bring this approach to large scale production level, which is why other strategies have attracted a renewed interest.

**3.2 Supported growth**

It has been known since the early 1970s that graphene could be grown directly on solid substrates and two different mechanisms can be exploited: the thermal decomposition of a carbon-rich faces of carbides or the epitaxial growth of graphene on metallic or metal carbide substrates by chemical vapor deposition of hydrocarbons. In Table 2, we compile the type of substrates, growth parameters of such techniques currently used in the direct growth of graphene or multi-layers graphitic samples as well as the specific structural characteristics of the graphene domains produced.

The thermal treatment of silicon carbide at about 1300°C under vacuum results in the sublimation of the silicon atoms while the carbon-enriched surface undergoes reorganization and, for high enough temperatures, graphitization.[28, 53, 67-69] The careful control of the sublimation has recently led to the formation of very thin graphene coatings over the entire surface of SiC wafers, with occasionally only one graphene layer being present.[54, 70] This growth method obviously raises many hopes that graphene could be more easily incorporated in the mainstream electronic industry. In particular, by confronting structural, spectroscopic and electronic data, it was initially thought that SiC-supported graphene was primarily monolayer although the lateral coherence length was limited to a few tens of nanometers,[54] which could be extended to about 200 nm by using the C-terminated faces.[71] However, Raman and STM studies have since shown that the symmetrical, but energy-shifted Raman signature is misleading and actually derives from the very loose stacking of the multilayers structures.[40, 72] In particular, the single Lorentzian Raman peak of epitaxial graphene strongly supports the observed quasi-two-dimensional Dirac-like character of the electronic states. Interestingly, the peak position is shifted by 15 to 25 $cm^{-1}$ above the 2640 $cm^{-1}$ peak of exfoliated graphene, indicating that a very weak and variable interlayer coupling results from a rotation of the layers with respect to each others. Graphitization of the carbon face of SiC under vacuum is accompanied by surface roughening and the formation of deep pits that induces the wide graphene thickness distribution and the limited lateral extension of crystallites. A major breakthrough for wafer-scale applications of epitaxial graphene has been the *ex-situ* graphitization of Si-terminated SiC(0001) in a 900 mbar argon atmosphere.[73] The annealing under controlled atmosphere promotes step bunching and the creation of wide and long insulating SiC terraces on which truly monolayered graphene is observed, except around the step edges.[53] The quality of graphene is also improved by a higher annealing temperature (1650°C) since sublimation of silicon occurs at 1500°C under argon rather than 1150°C in UHV. This method results in wafer-scale monolayered graphene with carrier mobility of 2000 $cm^2$/Vs at 27 K for a carrier density of ~ $10^{13}$ $cm^{-2}$, which is only five times less than exfoliated graphene on substrates in the limit of high doping. The prospects of large scale processing of graphene field-effect devices on SiC wafers, which initially showed modest performances compared to devices made from exfoliated graphene,[74, 75] are therefore reinforced.

Noteworthy, other carbides have been exploited early on to produce supported graphene. In particular, the decomposition of ethylene gas on the (100), (111) and (410) surfaces of titanium [76-78] and the (111) faces of tantalum [79, 80] carbides produces graphene monolayers. Interestingly the morphology of the TiC faces determines the graphene objects. In particular, large 200x200 nm monolayer crystallites are formed on terrace-free TiC (111), whereas monolayer graphene nanoribbons are obtained on the 0.886 nm-wide terraces of TiC(410).

The decomposition of hydrocarbons into graphitic materials has been extensively studied, in particular in the quest for a mass production method of carbon nanotubes.[81, 82] It is well documented that metal surfaces can efficiently catalyze this conversion and another route to supported growth of graphene consists in the chemical vapor deposition (CVD) on metallic surfaces.[55, 83] In an early method, the decomposition of ethylene on Pt(111) substrates at approximately 800 K forms nanometer-sized, uniformly distributed graphene islands. A continuous layer is formed at the lower step edges and additional large, regularly shaped islands are produced on the terraces upon further annealing above 1000 K. In the case of chemisorbed graphene, the size of the domains reflects directly the width of atomically flat terraces, which are a few nanometers for Pt(111). Recently, large single-crystalline graphene domains where grown by CVD of ethylene on Ru(0001).[56] Following a mechanism similar to the catalysis of carbon nanotubes from transition metal nanoparticles, ethylene is first dissolved in ruthenium at 1420 K. Supersaturation is then induced by lowering the temperature to 1100 K, which triggers the nucleation of graphene islands on Ru(0001) terraces.[84] The islands grow under the influence of the temperature gradient and form extremely large single crystalline domains that exceed 100 μm in lateral size. The first graphene monolayer is in strong interaction with the metal (carbon-metal bound is 1.45Å for Ru compared to 1.65Å for 4H-SiC(001)[85]) and the semi-circular morphology of the domain shows that up-going step edges stop the graphene growth, which can only progress towards down-going steps. In contrast, the second layer is weakly coupled to the metal because the first layer acts as an adhesion layer that efficiently screens off the residual charges from the underlying substrate. The moderate process temperature and large single crystal graphene obtained underline the relevance of metal-supported graphene production. However, a major drawback is the strong metal-graphene interaction, which limits the size, alters the transport study of pristine graphene and prevents device fabrication. Lateral extension can be improved in physisorbed systems by several approaches. For example when ethylene is replaced by benzene, a full coverage of the heated Pt substrate is observed at 1000 K and high precursor pressure.[83] On the other hand, the chemisorption of graphene on metal ensures the perfect epitaxy and hence the good structural quality of the graphene. This contradiction has been substantially lifted by resorting to iridium substrates, which produce a moderately chemisorbed graphene. Single crystalline layers are still produced but nevertheless spread over the up-going steps.[86] Low-pressure CVD of ethylene on Ir(111) between 1120 K and 1320 K results in the growth of graphene monolayer exhibiting full coherence over step edges on a few micrometer length scale. The extension of graphene-covered areas increases with increasing growth temperature and a further thermal annealing removes the few pentagon-heptagon edge dislocations observed. The observation of several micrometer-large Moiré patterns strongly suggests that structural coherence of the graphene lattice is maintained across atomic steps. This very promising combination of high quality epitaxial growth with a limited substrate-graphene cohesion has opened the prospects of growing and then transferring graphene onto any substrates including insulating or transparent ones. Iridium patterning and processing is rather limited but the equivalent CVD process has been shown for nickel, which is much more preferable for nanotechnological processes. Nickel is one of the best working catalysts for carbon nanotubes, which was quite naturally adapted to graphene growth.[87] Dresselhaus *et al.* have shown that polycrystalline nickel films could be used as a template for the CDV growth of centimeter-sized continuous graphene films composed of one to a dozen layer-thick graphene domains.[58] The lateral size of the graphene domains correlates directly with the nickel crystallite size, up to 20 μm, since the growth proceeds from a similar dissolution/precipitation mechanism to the ruthenium-templated method. The easy dissolution of nickel in dilute hydrochloric solution facilitates the transfer of graphene from the metal surface to a polymer coating and then from the polymer film onto any arbitrary substrate as illustrated in Figure 5.[58, 88] This approach, which has also been reported for copper,[89] demonstrates that transparent conductive graphene films can be produced in large areas without the limitation of the graphene oxide reduction step. With a mobility reaching 4000 cm$^2$/Vs, the large scale integration of devices can be envisioned by pre-patterning the nickel film with standard technologies.[58, 88] The metallic patterns are converted into identical graphene motifs, which are then transferred intact as illustrated in Figure 5. These recent proofs of principle indicate that methods will be rapidly available to incorporate graphene into low-cost, low performance devices. Yet, the supported growth methods are intrinsically bound to rather expensive templates which would be relevant only for high performance applications. To reach this goal, the current state-of-the–art will have to progress towards improved structural quality, better lateral resolution of template patterns and better preservation of electronic properties upon transfer printing.

For low performance applications, wet chemical approaches provide appealing alternatives capable of mass producing graphene showing satisfactory optical and electronic properties as depicted in the following paragraph.

### 3.3 Graphene oxide and the wet chemical routes

In order to extract graphene monolayers, micromechanical exfoliation relies on the balance between the interlayer cohesion and long distance interactions between the tape or the substrate and graphene. In contrast, chemical exfoliation proceeds by weakening the van der Waals cohesive force upon insertion of reactants in the interlayer space. Subsequently the loosened layer stacking is disrupted when the intercalant decomposition produces a high gas pressure. As a consequence, the $sp^2$ lattice is partially degraded into a $sp^2$-$sp^3$ sheet that possesses less π-π stacking stability. Chemical exfoliation can be performed in suspension and hence up-scalability is straightforward and it could offer a route to large scale graphene production.[90]

Historically, the oxidation of $sp^2$ graphite into less aromatic carbon has been known for more than 150 years after Brodie's work on the determination of carbon atomic mass and has only marginally evolved since then.[91-93] The oxidative intercalation of potassium chlorate in concentrated sulphuric and nitric acid is highly exothermic and produced modified graphite flakes composed of highly re-hybridized carbon sheets bearing hydroxyl and carboxyl moieties. This suspension was initially named graphitic acid but is now more commonly known as graphite oxide (GO). In 1958, a faster and safer route to graphite oxide was reported by Hummers,[94] where graphite is dispersed into a mixture of concentrated sulfuric acid, sodium nitrate and potassium permanganate (replacing the potassium chlorate) at 45°C for a couple of hours. The X-ray and electron diffraction investigation of graphite oxide show the total disappearance of the typical graphite inter-layer diffraction peak (0.34nm) and the appearance of a new one indicating a larger interlayer spacing (0.65-0.75nm) which strongly depends from the solution water content.[95] Graphite intercalation compounds obtained by the intercalation of sulfuric acid between the graphite layers are commonly used in chemical and electrochemical industrial applications, under the name of Expandable Graphite.[10, 96, 97] In this material, the graphite layers remain largely intact but the *d*-spacing is significantly enlarged so that the volume of the intercalated graphite is 100 times that of the starting natural graphite material. Expandable graphite is thus a readily available source for exfoliation and further colloidal dispersion of the resulting graphene oxide material.

After the intercalation of graphite by one of these three methods, few or even single-sheet materials are obtained by decomposing the intercalated reactant to produce a large amount of gas in the van der Waals space by chemical or thermal means. A rapid annealing to 1050 °C (at a heating rate exceeding 2000°C/min) generates a $CO_2$ over-pressure and splits the graphite oxide into individual sheets, which were observed by TEM as early as 1962.[44, 95, 98] Ultra-sonication can be used to further separate the graphite oxide into individual graphene oxide sheets. Graphene oxide solutions are yellow when high graphene oxide concentrations are present or greenish-blue when non-oxidized graphene sheets are the major constituents. The oxidative intercalation steps introduce phenol, carbonyl and epoxy moieties that account for the colloidal stability of graphene oxide suspension in polar solvents.[99, 100] The direct stabilization by electrostatic repulsion can be reinforced by mixing graphite oxide with cationic polyelectrolytes such as polyaniline, polyallylamine hydrochloride or aluminium Keggin ions,[101, 102] single-stranded DNA,[103] polystyrene sulfonate,[104] or polymers and surfactants such as phospholipid-polyethyleneglycol[105] or polystyrene.[106] The composite materials thus obtained are not only more stable in time but also easier to process, for example in thin films. In analogy to the nanotube bucky paper, thick graphene oxide paper can be produced by filtration to give composite sheets with exceptional tensile modulus.[107] Thinner deposits down to a few nanometers can be obtained by spin-coating or spraying the suspension in order to obtain a continuous transparent and conductive coating.[108-110] The directed deposition of graphene oxide flakes was demonstrated by Sheenan *et al.*, who took advantage of the negatively charged colloidal flakes to adsorb them onto gold-coated mica substrates which had been pre-patterned by micro-imprinting.[111] Tuning the solution pH, adsorption time, GO dispersion concentration and surface functionalization allowed the control of both the localization of the deposit and its thickness. Therefore, several opportunities exist to address the challenges of large area applications by these macroscopic graphene oxide films.

Graphene oxide bears interesting potentials as a single sheet platform for further chemical derivatization towards reinforcement material, sensor interface, etc., but if graphene oxide is to be considered as a large scale precursor of graphene, it is important to stress that these chemical approaches significantly modify the graphite layer structure and in particular converts a large fraction of the $sp^2$ carbon into $sp^3$ configuration. Graphene oxide shows fundamentally different electronic properties than individual semi-metallic graphene layer obtained by mechanical exfoliation or supported growth (See Section 4.1). A chemical reduction treatment is thus required to attempt a recovery of the specific properties of graphene. Once deposited on a substrate, the oxidized moieties of graphene oxide can be reduced by hydrazine vapors ($NH_2$-$NH_2$) or hydrogen plasma.[44, 109, 112, 113] The reduction by dimethylhydrazine results in the aggregation of the reduced flakes unless a stabilizing agent is simultaneously added to the suspension.[108, 114] In contrast, dried graphene oxide film are readily reduced and dispersed into pure hydrazine where it remains stable. [115]

Considering the notorious toxicity of hydrazine, an interesting green route to water-based graphene suspension consists in the quick deoxygenation of graphene oxide in alkaline conditions (0.1 M NaOH) at mild temperature (80°C).[116]

In spite of the recent efforts to pursue the chemical route to graphene, the reduction step remains incomplete, and the resulting material is an ill-defined intermediate between graphene and graphene oxide. Nevertheless, reduced graphene oxide is easily processed and, even for a partial recovery of $sp^2$ character, the sheet resistance is decreased by four to five orders of magnitude compared to the starting graphene oxide. The reasonable conductivity and the excellent transparency of reduced graphene oxide ultrathin films make it a promising component of transparent conductors as well as reinforcement and anti-aging agent in polymer composites as briefly summarized in Section 4 and detailed in reference [90].

While oxidative intercalation and exfoliation are the dominant approaches to the production of single sheet graphene derivatives, a softer method was reported by Coleman *et al.*, which is based on solvatation, *i.e.* the enthalpic stabilization of dispersed graphene flakes by solvent adsorption.[24] Inspired by the known variation in the efficiency of organic solvents to disperse purified nanotubes, the authors have directly dispersed and sonicated pristine graphite in solvents such as N-methyl-pyrrolidone (NMP), dimethylformamide (DMF), γ-butyrolactone (GBL) or dimethylacetamide (DMA). Although technically similar to the graphene oxide exfoliation, this simple protocol differs from it by the absence of the oxidation step. The dispersed material remains a concentrated suspension of defect free, un-oxidized $sp^2$ graphene, which can be further homogenized by centrifugation to yield solutions containing predominantly 1-μm diameter single sheets of graphene. Amongst all tried solvents, NMP appears to provide the best thermodynamic stabilization and so the most concentrated suspensions. Other groups have reported the successful dispersion of graphene monolayers in DMF [110] or benzene.[117] TEM imaging and electron diffraction, Infra-Red and Raman spectroscopy and X-ray photoelectron spectroscopy (XPS) data support the conclusion that solvent exfoliated carbon sheets are not oxidized and indeed genuine graphene. The dispersion in NMP can be further improved by intercalating graphite with alkaline metals such as potassium. In such intercalation compounds, the electron transfer from metallic potassium to graphene reduces the carbon atoms and yield negatively charged $sp^2$ graphene. Upon exposure to NMP, the sheets spontaneously exfoliate and form a stable suspension, even in the absence of ultrasonication.[118] Similar dispersions can be obtained in water provided that the solvent / graphene interface is stabilized by surfactant such as sodium dodecylbenzene sulphonate (SDBS). The colloidal suspension is less stable in water than in organic solvents. [119]

In oxidation or solvatation driven exfoliation, the separation of the layers relies on the mechanical disruption of the graphene stacking, usually by ultrasonication. When the ultrasounds are used extensively, not only are layers separated, but also fragmentation progressively occurs so that small graphene ribbons are produced and can be stabilized. Indeed, Dai *et al.* have shown that the thermal exfoliation of acid intercalated graphite and its subsequent dispersion in a 1,2-dichlorethane solution of poly(m-phenylenevinylene-co-2,5-dioctoxy-p-phenylenevinylene) by sonication for a prolonged time produced a solution with an appreciable amount of graphene nanoribbons with dimension ranging from 50 nm to sub 10 nm.[120] The ability to produce laterally confined ribbons is very promising since electronic properties of graphene are known to be governed by the edge states when the ribbon width becomes smaller than 10 nm (See Section 4.1). Such ribbons, and graphene itself, have often been figuratively presented as opened nanotubes, but the actual realization of such a transformation was reported recently by the groups of Dai and Tour. Indeed, the plasma etching of nanotubes that had been partially embedded in a polymer film [121] or alternatively the solution-based oxidative process to unfurl the nanotubes, followed by a chemical reduction step oxidized moieties [122] both provide 10-20 nm wide ribbons with smooth edges, a major improvement compared to lithographically defined and plasma etched ribbons. The recent demonstration that sub-10 nm ribbons were all semiconducting is an obvious asset of graphene nanoribbons (GNR) over nanotubes. [123]

Furthermore, these highly confined, aromatic objects are amenable to chemical functionalization while offering a continuous path through the graphene sheet to mesoscopic and even macroscopic dimensions. In this regards, mechanically or chemically exfoliated as well as template-grown graphene provide an unprecedented way to bridge physical properties at the nanoscale with single molecular phenomena. Interestingly, the opposite approach, namely the synthesis of graphene-like polyaromatic hydrocarbon has been developed over the past 15 years and is now reaching sizes that are comsparable to the narrowest nanofabricated ribbons thus bridging the entire size range from centimeters to one nanometer.

### 3.4 . A molecular approach

When graphene domain sizes exceed a few hundreds of nanometers, the observed behavior are well described by condensed matter models of mesoscopic phenomena. Yet graphene is intrinsically composed of the 2D pavement of organic molecular entities, benzene rings, which confer to graphene its Janus character, one face showing a mesoscopic 2D gas of $\pi$ electrons, the other face showing properties sensitive to the arrangement of molecular rings and edge atomic chains. In order to investigate the truly nanometer-scale regime, bottom-up approaches have been explored, which consists in starting from carbon precursors of small, yet well-defined, size and to convert them to graphene. One first strategy is the thermodynamic conversion of nano-diamond clusters into nanometer-sized graphene islands. In particular, Enoki and coworkers have shown that the electro-deposition of nano-diamond particles from an isopropanol-based solution onto a HOPG followed by a thermal graphitization process at 1600°C could transform the $sp^3$ precursors into $sp^2$ graphene islands. [124, 125] So far, mostly multilayer domains are obtained, which exhibit an interlayer distance of 0.35–0.37 nm in agreement with the graphite Bernal stacking. In contrast to nanoribbons prepared by prolonged ultrasonication of graphite [120] or even to the nanopatterning of graphene by ion damages [60, 126, 127] or by STM/AFM lithography,[128, 129] the island edges show a sharp zig-zag or armchair segments when imaged in STM.[130] These laterally confined nano-islands, with atomically precise edge structure, represent the upper form of molecular-scale graphene. Their study could offer the much sought-after intermediate size lying between mesoscopic and molecular description of graphene.

The bottom-up synthesis of ever larger polycyclic aromatic molecules has started over a century ago. Rules to determine their aromaticity, i.e. the molecular equivalent of the 2D π electron gas, have been discussed by Clar in 1972 and are now known as the aromatic sextet rule.[131] The synthetic routes towards nanoscale graphene have been largely explored by Müllen and co-workers who have produced a number of graphene precursors with specific structures that allow to rationally determine the morphology of the condensed oligomer. The core molecule is the hexabenzocoronene (HBC), which consists of thirteen fused hexagon aromatic rings (Figure 6a). Substituted HBC became itself a building block, along with other branched hexaphenylbenzene derivatives, for further oxidative cyclodehydrogenation reactions leading to planar graphene discs. In this way, graphene molecules containing as many as 90 [132] and up to 222 [133] carbon atoms in their cores and different substituents were obtained. The largest graphene molecule synthesized to date is shown in Figure 6b. Although molecules with such large molecular weights precipitate in solutions in spite of the side-chain substituents, individual molecule can still be manipulated and deposited on substrate by MALDI mass spectroscopy techniques.[134] The same method can be applied to one-dimensional fusion of polyaromatic precursor, which could be a rigorous approach to the total synthesis of graphene nanoribbons with controlled edge configuration (Figure 6d). [135, 136] A first step in this direction was the synthesis, deposition and characterization of dehydrogenated polyphenylene polymers which yielded 2.9 nm-wide ribbons having length of 8 to 12 nm (Figure 6c).[137] The electronic properties of these polyaromatic have been characterized by STM studies only,[138] but the design of the polyaromatic core and the peripheral solubilising side chains already allows these materials to be prepared as thin films showing liquid crystal properties. Hence molecular graphene could rapidly be one other route to graphene-based optical devices.

**4. The versatile properties of graphene.**

The rapid adoption of graphene as a material of interest lies in its actual availability by the range of techniques and methods described before, but also, and maybe principally, because monolayer and few layer graphene and graphene oxide possess a diverse set of unusual properties. These properties happen to be matching the short-comings of other materials, such as carbon nanotubes, graphite, heterostructure 2D electron gas or indium tin oxide, that have been studied and used for some time. In this regard, several communities try to investigate and exploit graphene properties to circumvent these current roadblocks. The dynamism of the graphene research is thus rooted in the convergence of expectations, hopes and curiosity, which are amplified by the awesome simplicity of the single atom thick sheet. In the following paragraphs, we will not give an exhaustive account of all observed properties and phenomena since focused review articles are readily available, but, in regards to the production methods described above, we will show how graphene is opening new horizons.

**4.1 Electronics**

The remarkably high electron mobility in graphene at room temperature, μ, which exceeds 2 000 cm$^2$/V.s for any micromechanically deposited graphene, has offered an immediate access to an unusual quantum Hall effect.[139, 140] Hence from the first reports, the mesoscopic physics community was offered a new object in which many gedanken experiments could be implemented experimentally. Furthermore, the simultaneous observations of high mobility, sensitivity to field effect and large lateral extension, which implied an ease of contacting, turned graphene into an appealing alternative to carbon nanotubes for field effect transistors devices.[2, 16] Graphene mobility is temperature-independent between 10 and 100 K, suggesting that the dominant scattering mechanism was initially related to graphene defects.[139, 141] Improved sample preparation, including post-fabrication desorption of adsorbates by current annealing, allowed to reach mobilities of more than 25 000 $cm^2$/V.s. For oxide-supported graphene devices, trapped charges are the dominant scatterers [142] but optical phonons of the substrate are setting the fundamental upper limit of 40 000 $cm^2$/V.s at room temperature.[143] Further enhancing the mobility of graphene thus requires an efficient screening[144] or even the complete removal of the substrate. Indeed, in suspended and annealed graphene devices, the mobility has been shown to exceed 200 000 $cm^2$/V.s (Figure 7). [62, 145, 146] This is the largest ever reported value for a semiconducting metal. In addition, the mobility remains high even at the highest electric-field-induced carrier concentrations, and seems to be little affected by chemical doping,[147] suggesting the possibility of achieving ballistic transport regime on a sub-micrometer scale at 300 K.[148] Moreover, experimentally measured conductance indicates that the value of the mobility for holes and electrons is nearly the same.[149] Graphene is characterized by a minimum value of the conductance, which scales as $e^2$/h, even in the limit of zero-charge density, both from a theoretical [150] and experimental [142] point of view. From conductivity measurements, the mean free path, $l_e$ can be estimated. In a suspended sample before and after annealing, $l_e$ increases from ~150 nm to ~1 μm, the latter distance being comparable or even larger than the characteristic size of nanofabricated devices as shown in Figure 7.[145] Electronic transport in such devices where no defects, neither from the substrate nor from an unclean sample, interfere with the carrier conduction is no longer diffusive, but reaches the ballistic limit.[151, 152]

The second unusual feature in electronic properties of graphene is the almost strict linear dispersion curve near the Dirac point where the Fermi level can easily be set by adjusting the doping level electrostatically (Figure 2b). This resulted in the successive discovery of a Berry's phase or a Quantum Hall effect (QHE) with fourthly degenerated Landau levels due to K and K' valley and pseudo-spin degeneracy except for the zero Landau level which is only twice degenerated due to strict electron-hole symmetry near the Dirac point.[8, 139, 140, 153-155] Similarly, bilayer graphene exhibits a distinct behaviour resulting from the zero-gap parabolic dispersion with two extra bands available.[141, 156] Among the remarkable quantum electrodynamic properties of graphene, the room temperature QHE,[157] 1/3 fractional QHE,[158, 159] Klein tunneling paradox, [160-162] and the breakdown of the Born-Oppenheimer approximation [163] have been observed. For a detailed account of these specific topics, the Reader is invited to refer to the excellent recent review articles dedicated to these aspects of graphene.[7, 8, 154, 164-166] Along with the large electronic coherence length, spin coherence length is also large enough (~ 1 μm) to allow for the fabrication of spintronics devices [167-169] and the observation of Andreev reflections.[161, 170-173] Spin precession over micrometer scale distances has also been demonstrated in single graphene sheets at room temperature.[174] However, device fabrication usually exploits non-linear effects since information processing requires signal amplification for improved detectability. The linear dispersion curve with vanishing density of states at the meeting point of the valence and conduction bands forbids such non-linearity. In particular, field-effect induced modulation of transport characteristics is quite inefficient in graphene. This prompted the quest for engineering an electronic gap in graphene, which would restore the capability to use graphene in field-effect transistor (FET) devices.[175]

Several strategies are available to open a gap in graphene for FET device fabrication. Among them, one can cite: (i) the inclusion of $sp^3$ hydrocarbon defects in the $sp^2$ lattice and (ii) the distortion of the carbon atom lattice under uniaxial strain.[176-178] The former approach has been recently illustrated by the partial hydrogenation of graphene [179, 180] towards the fully hydrogenated, $sp^3$ graphene-derived carbon sheet called graphane.[181] Meanwhile, the most straightforward approach is the basal interaction of graphene with a substrate.[182] When the substrate is composed of graphene sheets, as in graphite, the observed gap of the uppermost layer is typically a few tens of meV, which is too weak for gaining substantial conductivity modulation. The strong chemisorption of graphene grown on metals such as nickel and ruthenium induces an expected large (~1-2 eV) gap in contrast to physisorbed growth on iridium, for example,[183] but applications of metal-templated graphene are limited. Similarly, the influence of the SiC substrate in epitaxial graphene grown on the Si-rich face has been shown to induce a gap as large as 260 meV by ARPES.[32, 184] The gap is partially modulated by lateral size confinement,[185-187] but infortunately, upon device fabrication, the effective gap is almost negligible probably due to the alteration of the substrate-graphene interface.[188]

The earliest proposition that bandgap could be induced in semi-metallic graphene was a direct consequence of the lateral confinement of the 2DEG in ribbon-shaped graphene.[9, 189-192] It has thus been predicted that the electrical properties of graphene nanoribbons can be tuned from perfectly metallic, for zig-zag edge ribbons (ZGNR), to semiconducting behavior, for armchair ribbons (AGNR). In this later case, the gap varies with the ribbon width,[193] length and topology in analogy to carbon nanotubes, the chirality being replaced by the edge morphology.[194, 195] Experimentally, graphene nanoribbons with sub-100 nm width have been first produced by standard lithographic patterning (with the inorganic resist hydrogen silsesquioxane, HSQ) combined with oxidative reactive ion etching in order to locally remove the carbon monolayer.[126, 127, 196] Alternatively, graphene could be etched directly by using focused ion beam (FIB).[60] In both approaches, non-linear transport characteristics have been observed, which can be modulated by global or local gates applied either perpendicularly or transversely to the ribbon (Figure 8). For sub-100 nm width, W, a clear gap is observed in the I-V characteristics the energy value, $E_G$, of which varies as $W^{-1}$ (Figures 8a,b) This variation has been interpreted as a direct evidence of the lateral confinement of the 2DEG.[126, 197] Alternatively, the roughness of the imperfect edges that should be expected from ion damage patterning techniques, as well as the residual yet probable partial amorphization of the ribbon upon exposure to direct or back-scattered ions, can be considered as a source of localization. Therefore, non-linear conduction through ion-etched ribbons can be understood as tunneling through a unidimensional series of graphene dots separated by highly resistive barriers (Figure 8c).[60, 196, 198] In contrast, the FET devices based on the 50 to 5 nm wide nanoribbons obtained by the Dai's method by ultrasonication and polymer stabilization,[120] show ambipolar transconductance characteristics with $I_{ON}/I_{OFF}$ ratio reaching as much as $10^6$ (Figure 8d). Interestingly, all sub-10 nm ribbons were semiconducting at room temperature with mobilities limited to 200 cm$^2$/V.s.[123] If this production method obviates irradiation damages and thus preserves a good quality graphene during fabrication, the edges are most certainly rough enough to degrade the graphene band diagram and ensure the observed semiconducting nature of the nanoribbons. One other noteworthy point is that similarly to carbon nanotubes, Shottky barrier are observed at the metal / graphene interface when individual sub-10 nm wide ribbons are connected. Although the work function adjustment is now well mastered, this pleads in favor of graphene nanoribbon FET that incorporates larger lead electrodes in the same graphene sheet. These early attempts in engineering the electronic gap of high quality graphene by lateral confinement point to the extremely challenging need to perform atomically precise tailoring of graphene over long distances to obtained high quality ribbons for advanced graphene electronic architectures.

In addition to these high added value applications, thin films of reduced graphene oxide or solvent exfoliated graphene as well as transferred CVD grown graphene possess high enough a conductivity and an optical transparency to be considered as a viable competitor to the industrial standard, Indium Tin Oxide (ITO). ITO has a typical sheet resistance of 40 Ω/sq for visible transparency above 80%. As-produced graphene oxide (GO) shows insulating transport characteristics but conductivity can be improved by reduction of the *sp$^3$* carbon atoms in solution. The casting of suspended reduced GO flakes into sub-3 nm thick films yield >90% transparent coating with sheet resistance of the order of $10^9$ Ω/sq which can be reduced to ca. 40 kΩ/sq upon thermal annealing.[108, 109, 112] Thin films obtained from solvent exfoliated graphene have a sheet resistance of about 6 kΩ/sq which can be further reduced by chemical doping to 50 Ω/sq.[110] Finally the transfer of graphene grown by CVD on nickel from the metallic template to a insulating substrate yield large area films of few layer graphene with sheet resistance of about 280 Ω/sq for a transparency in the visible range beyond 80%.[88]

These selected examples clearly illustrate how the remarkable electrical properties of graphene make this material highly relevant for both high mobility or even ballistic transport applications as well as low cost materials for transparent conductive materials such as liquid crystal cell electrode. This particular aspect takes advantage of electronic as well as optical properties of single sheet graphene, the latter ones being discussed next.

**4.2 Optics**

Black graphite soot or silverish graphite crystal actually turns into highly transparent when thinned down to a graphene monolayer as illustrated in Figure 9. This optical characteristic combined with the excellent conductivity of graphene-based materials holds promises as a replacement to the cost-raising standard ITO. Indeed, in the visible range, thin graphene films have a transparency that decreases linearly with the film thickness (Figure 9b). For 2-nm thick films, the transmittance is higher than 95% and remains above 70% for 10-nm thick films.[110, 199] Moreover, the optical spectrum is quite flat between 500 and 3000 nm with the dominant absorption being below 400 nm. [24] Considering the low cost of chemically exfoliated graphene compared to ITO (40 Ω/sq at > 80% transmittance) or carbon nanotube mats (70 Ω/sq at 80% transmittance), the combination of high film conductivity, optical transparency, chemical and mechanical stability immediately suggests employing graphene as transparent electrode for solar cells or liquid crystal but also as processable transparent flexible electrode material.[110, 199, 200]

Surprisingly, the macroscopic linear dependence of the transmittance with the thickness of graphene films is intimately related to the two-dimensional gapless electronic structure of graphene. When one and two layered graphene sheets are suspended, the measured white light absorbance is 2.3 and 4.6% respectively with a negligible reflectance (< 0.1%) (Figure 9a).[201] The strict linearity has been further demonstrated up to five monolayers (Figure 9a, inset) and the spectral variation of the absorbance matches the one observed for macroscopic thin films. One can demonstrate that the transparency of graphene depends only on the fine structure constant $\alpha = 2\pi e^2/hc = 1/137$ (*c*, speed of light), which describes the coupling between light and relativistic electrons, typically associated with quantum electrodynamics phenomena rather than materials science. Within this framework of ideal Dirac fermions, the dynamic conductivity of graphene in the visible range, G, is found to depend only on universal constants: $G \sim \pi e^2/2h$. The absorbance of few layer graphene is simply the product of $\pi\alpha$ by the number of layers.

Further, in the Near-InfraRed (NIR) to Mid-InfraRed (MIR) range, properties of graphene offer the possibility to investigate the intimate connection between the energy dispersion relation of electrons, holes, and plasmons, the latter playing an important role in the intraband adsorption in the THz ($10^{12}$ Hz) range. In the high energy limit, optical absorption increases with increasing energy, suggesting possible deviation from the Dirac fermions dispersion relation for energy greater than 0.5 eV due to trigonal warping. Furthermore, the small group velocity of plasmons together with a strongly confined field, suggests the possibility of obtaining large value of gain in stimulation-emission phenomena at optical and infrared frequencies, allowing for the fabrication of graphene-based THz emitters, plasmon oscillator and non-linear optical device.[202, 203]

Transport properties in graphene can also be probed by infra-red spectroscopy because of active interband transitions. Measuring IR reflectivity changes while applying a gate voltage showed that these interband transitions in graphene could be modulated. It confirmed the linear dispersion of Dirac fermions in graphene monolayer and, in the case of bilayer, the strong dependence of the transitions from the interlayer coupling revealed van Hove singularities.[204] When a magnetic field, B, is applied, infrared magneto-transmission experiment can probe transitions between Landau levels.[205, 206] The absorption band variations as $B^{1/2}$ confirm the 2D Dirac fermions nature of the carriers and provide an independent measurement of the Fermi velocity. Insight on the carrier dynamics and relative relaxation timescales in graphene are provided by ultrafast optical pump-probe spectroscopy, which identifies two separate time regimes: an initial fast relaxation transient (tens of fs) is followed by a slower relaxation process (~0.4-1.7 ps). Those two regimes are related to carrier-carrier intraband and carrier-phonon interband scattering processes in graphene, respectively.[207]

Again in optical physics, graphene presents a variety of specific behaviour which spans application potentials in material science as well as fundamental transport and its ideal structure offers a quite direct and insightful link between both aspects.

**4.3 Mechanics**

Graphite, diamond and carbon nanotubes have each set their own record in terms of mechanical robustness, be it hardness or Young's modulus. Graphene is no exception although its mechanical behaviour has been much less investigated than its electronic or optical properties. Figures 10a and 10b illustrate how AFM nanoindentation was used to probe the elastic stress-strain response. The reported stiffness of the order of 300-400 N/m, with a breaking strength of ca. 42 N/m, represents the intrinsic strength of a defect-free sheet.[208] The estimates of the Young's modulus yielded approximately 0.5-1.0 TPa which is very close to the accepted value for bulk graphite.[208, 209] Interestingly, in spite of their defect, suspended graphene oxide sheets retained almost intact mechanical performances with a Young's modulus of 0.25 TPa.[210] These values combined with the relative low cost of thin graphite and the ease of blending graphene oxide into matrices,[114] makes these materials ideal candidate for mechanical reinforcement.[211-213] On the other hand, with such a high sustainable tension in a single sheet, graphene bears tremendous potential as the ultimately thin material for NEMS applications such as pressure sensors, and resonators. Mechanically exfoliated single- and multilayer graphene sheets placed over trenches in a $SiO_2$ substrate were contacted to produce graphene-based nano-electromechanical systems (Figures 10c,d).[214] Fundamental resonant frequencies, which can be triggered either optically or electrically, were experimentally observed in the 50-200 MHz range with a room temperature charge sensitivity as low as $8\times10^{-4} e/Hz^{1/2}$ and a quality factor in vacuum ($< 10^{-6}$ torr) of 80. Direct imaging of the driven oscillating modes by a non-conventional AFM technique revealed that non-uniformity of the initial stress upon depositing the graphene sheet result in exotic nanoscale vibration eigenmodes with maximal amplitude on the graphene sheet edges.[215] Suspended graphene being under tension and impermeable,[216] it provides the optimized atomically thin supporting membrane for sensitive gas sensing.

**4.4 Graphene sensors**

Sensing is a complex function which requires the integration of a number of properties from interface accessibility to transduction efficiency, molecular sensitivity and mechanical or electrical robustness. It appears that graphene fulfills many of these requirements and the implementation of highly sensitive graphene sensors could illustrate how the previously described properties can be brought in synergy to this specific objective. Indeed, suspended graphene is a pure interface with all atoms exposed to the environment. It is chemically stable yet functionalization towards specific interaction is possible. Moreover, local destruction of the $sp^2$ lattice preserves its mechanical robustness and does not jeopardize its 2D delocalized transport properties unlike carbon nanotubes. Electronic and mechanical properties can be exploited to perform the transduction of the sensing signal. Lastly, graphene and its oxidized forms are produced at a much lesser expense than other graphitic materials, such as nanotubes, currently considered for sensing platforms.

In the event of a gas molecule adsorbing onto graphene surface, the local change in the carrier concentration induces a doping of the delocalized 2DEG, which can be monitored electrically in a transistor-like configuration.[217] In contrast to other materials, the high mobility, large area ohmic contact and metallic conductivity observed in graphene contribute to limit the background noise in transport experiment. This confers to simple graphene FET device a high enough sensitivity to detect parts-per-billon levels or even single molecular events at a rapid rate, which has been realized experimentally.[147] Interestingly, cleaned graphene (i.e. typically after a treatment to remove residual lithography resist) showed a much attenuated sensitivity.[218] Further, functionalization of graphene, or even better reduced graphene oxide, is thus needed to improve sensitivity, chemical affinity and selectivity.[218-221] Graphene sensitivity is not necessarily limited to chemical species, but it can be generally applied to any phenomena capable of inducing a local change in the carrier concentration, such as presence of magnetic field, mechanical deformation or external charges.

An alternative to electrical detection scheme is to exploit the electromechanical behaviour of suspended graphene in analogy to carbon nanotube nanomechanical sensors that have been demonstrated to reach single atom sensitivity.[222] The large area and stiffness of suspended graphene and its very specific Raman signature plead in favor of developing mechanical mass sensors.[223]

**5. Conclusion: Graphene in the post-Silicon and post-CMOS eras.**

In reviewing recent work, we have been driven to consider the material properties of graphene as well as its peculiar mesoscopic behaviour, in particular regarding its electronic properties. The accumulation of non-conventional properties has attracted the attention of the microelectronics and the carbon nanotube communities which are now actively developing technological approaches to graphene-based transistors. It is remarkable how graphene has rapidly passed through the traditional stages between a new material under the scrutiny of scientists and the exploitation of its properties in application-driven devices in the hands of technologists. Now recognized as a promising actor in the post-Silicon era, graphene performances as transistor channel and the compatibility of its production with existing industrial microelectronic processes are being actively assessed in single gate, top-bottom gates and double side-gates configurations. Experimentally, the actual mobility in graphene channel after processing remains significantly higher than silicon and SOI (Silicon on Insulator) equivalent devices, even with a top gate electrode. [224, 225] From non-equilibrium Green's function transport simulations in the short channel limit, i.e. below 10 nm, the small effective mass of carrier in graphene facilitates strong source-drain tunnelling. Therefore the minimal leakage current should significantly increase below 10 nm so that graphene FET might not allow to extend the ultimate scaling limit of Si MOSFETs.[226] However the intrinsic switching speed of a graphene nanoribbon Schottky barrier FET, is found numerically to be several times faster than that of Si MOSFETs, which could lead to promising THz electronics applications. [226, 227] So far, operation of graphene transistors at GHz frequencies has been demonstrated.[228] Finally, when compared to carbon nanotube FET (CNT-FET) graphene transistors are easier to produce with CMOS processes compatibility. Electrode contacts are better and current densities are potentially much higher than parallel nanotube CNT-FET. In this respect, one can anticipate a rapid assessment of the actual relevance of graphene for post-silicon CMOS technology.

Yet large enough $I_{ON}/I_{OFF}$ will require the fabrication of very narrow ribbons with atomically precise edges.[229] Transmission electron microscopy (TEM) has recently proven that unsupported zig-zag and armchair ribbon edges were stable in ambient conditions in vacuum. [230, 231] Moreover, upon electron beam irradiation or current annealing, these edges are reconstructed with zig-zag edges showing a better stability. If atomic fabrication resolution has not yet been achieved, it is very promising that the observed edges are definitely atomically sharp. Similarly Scanning Tunneling Microscopy (STM) has characterized stable graphene edges of supported flakes.[232] STM and AFM oxidation of graphene on conductive or insulating substrates respectively is being developed as direct, resist-free method to pattern graphene down to the atomic scale.[128, 129, 233, 234] Finally, the nanoparticle-catalyzed cutting of graphene shows a high dependency on crystallographic orientation. Linear trenches of several micrometers are tus produced as the active nanoparticles travel along graphene crystallographic direction during the oxidation reaction.[235-237] If one controls the nanoparticle size and trajectory [238] as well as the oxidation reaction rate, one can anticipate that large scale cutting of graphite with extremely high resolution will be achieved.

With characterization and patterning of graphene reaching the atomic scale, graphene could offer another unique opportunity: the opportunity to bridge the gap with molecular electronics. This would open a way into a post-CMOS area, since CMOS architecture has no specific justification at atomic scale.

This last example, as many others examined all along this Review, illustrates that graphene possesses a combination of special properties which happen to answer several limitations of currently known materials and systems. If anything else, graphene stimulates an intense and exciting research activity and gets several scientific communities to dream of new concepts and to thrive to make them real.

**Acknowledgements**

The authors are grateful to C. Joachim, A. Loiseau, S. Roche and M. Monthioux for their support. E.D. would like to thank J.-F. Dayen, S. Linas, R. Diaz for continued efforts and valuable discussions. The authors acknowledge the financial support from the Agence Nationale de la Recherche (Contracts ANR-JC05-46117 and ANR-06-NANO-019) and the European Research Council (Contract ERC-2007-StG-203672). AM thanks the Higher Education Commission of Pakistan.

**FIGURE LEGENDS**

**Figure 1.** Low-dimensional carbon allotropes: fullerene (0D), carbon nanotube (1D) and graphene (2D).

**Figure 2.** Atomic and electronic structures of graphene. (a) Graphene lattice consists of two interpenetrating triangular sub-lattices, each with different colors. The atoms at the sites of one sub-lattice, (i.e. A) are at the centers of the triangles defined by the other lattice (i.e. B), with a carbon-to-carbon inter-atomic length of 1.42 Å. (b) $\pi$-$\pi$* band structure of graphene. The three dimensional first Brillouin zone is displayed in red and blue for the valence and conduction $\pi$ bands respectively above the planar projection of the valence band. The six Dirac cones are positioned on a hexagonal lattice. (a) Reproduced with permission from [7]. Copyright 2007 AIP.

**Figure 3.** Major techniques for graphene characterization. (a) Optical microscopy, (b) Atomic Force Microscopy (AFM), (c) Transmission Electron Microscopy (TEM), (d) Angle-Resolved Photoemission Spectroscopy (ARPES), (e) Raman scattering and (r) Rayleigh scattering. Adapted with permission from (c) [23], copyright 2008 American Chemical Society; (d) [27], MacMillan Publishers Ltd: Nature Physics, copyright (2006); (e) [41], copyright 2007 American Chemical Society; (f) [42], copyright 2007 American Chemical Society.

**Figure 4.** Micromechanically exfoliated graphene. Optical images of (a) thin graphite an d (b) few layer graphene (FLG) and single layer graphene (lighter purple contrast) on a ~300 nm $SiO_2$ layer. Yellow-like color indicates thicker samples (~100s of nm) while bluish and lighter contrast indicates thinner samples.

**Figure 5.** Nickel-grown graphene. (a) Optical image of a pre-patterned Ni film on $SiO_2$/Si. CVD graphene is grown on the surface of the Ni pattern. (b) Optical image of the grown graphene transferred intact from the Ni surface in (a) to another $SiO_2$/Si substrate. Reproduced with permission from [58]. Copyright 2009 American Chemical Society.

**Figure 6.** Molecular graphene. (a) Structure of the hexabenzocoronene (HBC). (b) Structure of the largest polycyclic aromatic synthesized to date, which contains 222 carbon atoms. (c) STM image at the solid/liquid interface on HOPG of self-assembled graphene nanoribbons shown in (d). Reproduced with permission from (a,b)[138], copyright 2007 American Chemical Society; (c,d) [137], copyright 2008 American Chemical Society.

**Figure 7.** Carrier mobility in graphene. (a) Conductance of a suspended graphene sample (—, blue) before and (—, red) after annealing as a function of carrier density, n, at a temperature of 40 K. The black dotted line is the expected behavior for ballistic transport. (b) Mobility displayed as a function of n exceeds 200 000 $cm^2$/V.s at n = $2\times10^{11}$ $cm^{-2}$. Reproduced with permission from (a) [145], copyright 2008 APS and (b) [62], copyright (2008) from Elsevier.

**Figure 8.** Graphene nanoribbon (GNR) field effect devices. (a) Apparent gap energy variation as a function of nominal GNR width (W < 100 nm) in backgated field effect devices obtained by lithography and ion etching. Inset: SEM image of a series of parallel GNR of different width. (b) Temperature dependence measurements of the transfer characteristics (source-drain current, $I_{ds}$, vs backgate voltage, $V_{gs}$) in a 20-nm wide GNR field effect device at source-drain voltage fixed at $V_{ds}$ = - 1mV. Temperatures are 400 K (black), 200 K (blue), 100 K (green), 50 K (orange), 4 K (red). The minimum current through the nanoribbon decreases by more than 1.5 orders of magnitude. Inset: False-colored SEM image of GNR devices fabricated by lithography and ion etching. The GNR widths from top to bottom are 20, 30, 40, 50, 10 and 200 nm. (c) Source-drain I-V characteristics at 40 K of a 50x2000 nm double side-gated nanoribbon with side gate voltage varying between 0 V (purple) and 30 V (dark red). Inset: False-colored SEM image of the few-layer graphene nanoribbon device with graphene source, drain and ribbon (cyan) as well as side gates (blue) etched by Focused Ion Beam. (d) Room-temperature transfer characteristics (current versus gate voltage $I_{ds}$-$V_{gs}$) for a 9 nm wide and 130 nm long bilayered GNR field effect device with Pd contacts and Si backgate. Inset: AFM image of the device. Scale bar is 100 nm. Reproduced with permission from (a) [126], copyright 2007 APS; (b) [197], copyright (2007) from Elsevier Science; (c) [60], Copyright Wiley-VCH Verlag GmbH & Co. KGaA; (d) [120], copyright 2008 AAAS.

**Figure 9.** Optical transparency. (a) Optical micrograph of one- and twp-atom thick graphene crystals. The optical transmittance at 550 nm as a function of the lateral coordinate shows clear steps, the height of which is proportional to the hyperfine constant. In inset, the linear variation of the transparency as function of number of layers is observed up to six layers. (b) Transmittance at λ~550 nm as a function of the thickness of reduced GO thin films, assessed indirectly by the total volume of filtered suspension. Plots are shown for thin films with different reduction steps. (a) From [201]. Reprinted with permission from AAAS; (b) Reproduced with permission from [109], MacMillan Publishers Ltd: Nature Nanotechnology, copyright (2008).

**Figure 10** Mechanical properties. (a) Schematic of AFM nanoindentation on suspended graphene. (b) Loading/unloading curve for device in (a) with modeling comparison (red line). (c) Schematic of a suspended graphene resonator. (d) Amplitude *vs.* frequency resonance curve taken with optical drive for the fundamental mode of a single-layer graphene resonator. (a) From [208]; (b) from [214]. Reprinted with permission from AAAS.

**Table 1** Advantages and disadvantages for techniques currently used to produce graphene.

**Table 2** Synoptic table of the graphene growth methods on both metals and carbides substrates. Acronyms are as follows: L, single layer coverage; ML, monolayer; T, temperature; STM, Scanning Tunneling Microscopy; LEED, Low-Energy Electron Diffraction; AES, Auger Electron Spectroscopy; AFM, Atomic Force Microscopy; PCM, Point-Contact Microscopy; XPS, X-Ray Photoemission Spectroscopy; ARUPS, Angle-Resolved Ultraviolet Photo-Electron Spectroscopy; SEELFS, Surface Extended-Energy-Loss Fine-Structure; SEM, Scanning Electron Microscopy; XPD, X-Ray Photoelectron Diffraction; TDS, Time-Domain Spectroscopy; LEIS, Low-Energy Ion Spectroscopy.

|  | Advantage | Disadvantages |
| --- | --- | --- |
| **Mechanical Exfoliation** | Low cost and easy<br>No special equipment needed,<br>$SiO_2$ thickness is tuned for better contrast | Serendipitous<br>Uneven films<br>Labor intensive (not suitable for large-scale production) |
| **Epitaxial Growth** | Most even films (of any method)<br>Large scale area | Difficult control of morphology and sorption energy<br>High temperature process |
| **Graphene Oxide** | Straightforward up-scaling<br>Versatile handling of the suspension<br>Rapid process | Fragile stability of the colloidal dispersion<br>Reduction to graphene is only partial |

**Table 1**

**Table 2** Advantages and disadvantages for techniques currently used to produce graphene.

| Substrate | | Growth Condition (gas, T, exposure)* | Experimental Technique | Edges | Comment ($a_C$=0.245nm) | Ref. |
|---|---|---|---|---|---|---|
| **METALS** | Pt(111) | Benzene ($C_6H_6$), T=1000K. 1-5L → non-graphitic film > 5L → full coverage | STM, LEED, AES | Hexagonal arrangement beyond edges | | [83] |
| | | Ethylene ($C_2H_4$), T=800K. 5L exposure (If T>1000K →graphitic island) | LEED, STM | No clear hexagonal arrangement; No growth over the edges | $a_{Pt}$ = 0.278nm $a_{Moire'}$ = 2.2nm | [55] |
| | | HOPG on 1ML graphitic film | AFM, PCM | Continuous film from upper terrace to lower terrace | 0.738nm < a < 2.1nm | [239] |
| | Pt(755) | Chemical Vapor Deposition | LEED, XPS, ARUPS | Formation of large sheet | | [240] |
| | Ni(111) | Chemical Vapor Deposition | LEED, AES Vibrational spectro | | Evidence of Fuchs–Kliewer phonons | [87] |
| | Ni(110) | Carbon monoxide (CO), T=600K. 90000L exposure | SEELFS | | Graphitic layer on [110] faces | [241] |
| | Ru(001) | Ethylene ($C_2H_4$), T=1270K. T-dependent solubility gradient | LEEM, SEM, μ-Raman, AES, electrical | No growth "uphill" over the edges | $a_{Ru}$=0.271nm a=0.145nm (1st layer) $a_{Moire'}$= 3nm | [56] |
| | Ir(111) | Ethylene ($C_2H_4$), T>1100K. 1L exposure | STM | Growth beyond both side of the edges | $a_{Ir}$=0.272nm $a_{Moire'}$= 2.5nm | [86] |
| | Co(0001) | Acetylene ($C_2H_2$), T=410K. 0.6L-3.6L exposure | XPS, XPD, LEED, TDS, LEIS | | K enhances the coverage of the surface | [242] |
| **CARBIDES** | nH-SiC (n=1,2,..) | Si sublimation, T~1670K. | LEED, X-Ray, STM | Formation of large continuous sheet over terraces | | [69] |
| | TiC(111) | Chemical Vapor Deposition on faceted surface, T=1770K | XPS, ARUPS, LEED | No edge-localized state | Growth on each facet | [78] |
| | TiC(410) | Chemical Vapor Deposition on platelets surface, T=1770K | XPS, ARUPS, LEED | No growth over the edges | Nanoribbon growth (~.1-2 nm) | [78] |
| | TaC(111) | Ethylene ($C_2H_4$), T=1570K 10000L exposure, T=1270K | AES, LEED, STM | Coverage is interrupted at terrace interface | a=0.249nm (1st layer) a=0.247nm (2nd layer) | [80] |

**Table 2**

**Table 2** Synoptic table of the graphene growth methods on both metals and carbides substrates. Acronyms are as follows: L, single layer coverage; ML, monolayer; T, temperature; STM, Scanning Tunneling Microscopy; LEED, Low-Energy Electron Diffraction; AES, Auger Electron Spectroscopy; AFM, Atomic Force Microscopy; PCM, Point-Contact Microscopy; XPS, X-Ray Photoemission Spectroscopy; ARUPS, Angle-Resolved Ultraviolet Photo-Electron Spectroscopy; SEELFS, Surface Extended-Energy-Loss Fine-Structure; SEM, Scanning Electron Microscopy; XPD, X-Ray Photoelectron Diffraction; TDS, Time-Domain Spectroscopy; LEIS, Low-Energy Ion Spectroscopy.

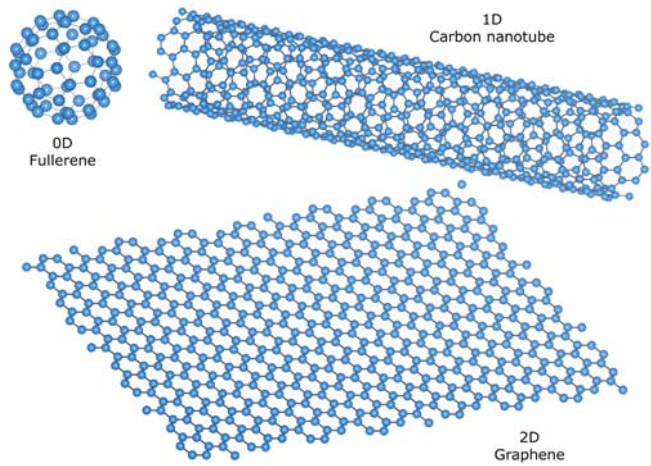

**Figure 1**

**Figure 1.** Low-dimensional carbon allotropes: fullerene (0D), carbon nanotube (1D) and graphene (2D).

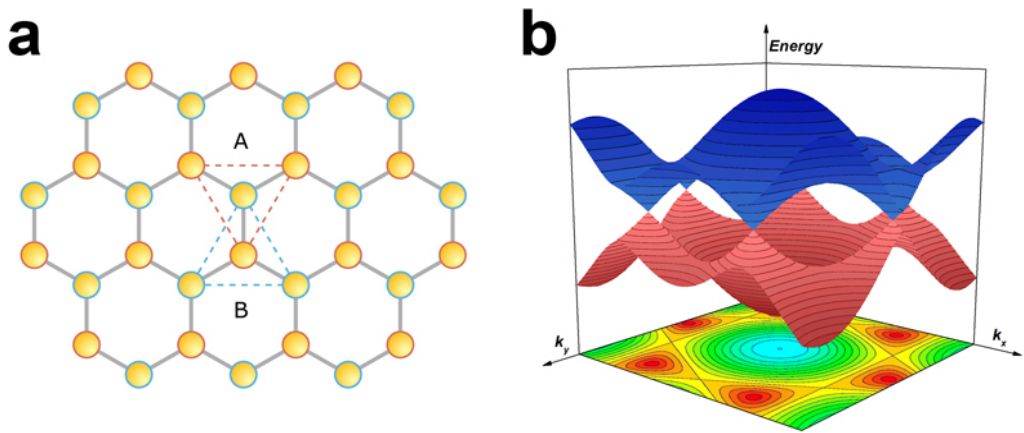

**Figure 2**

**Figure 2.** Atomic and electronic structures of graphene. (a) Graphene lattice consists of two interpenetrating triangular sub-lattices, each with different colors. The atoms at the sites of one sub-lattice, (i.e. A) are at the centers of the triangles defined by the other lattice(i.e. B), with a carbon-to-carbon inter-atomic length of 1.42 Å. (b) $\pi$-$\pi$* band structure of graphene. The three dimensional first Brillouin zone is displayed in red and blue for the valence and conduction $\pi$ bands respectively above the planar projection of the valence band. The six Dirac cones are positioned on a hexagonal lattice. (a) Reproduced with permission from [7]. Copyright 2007 AIP.

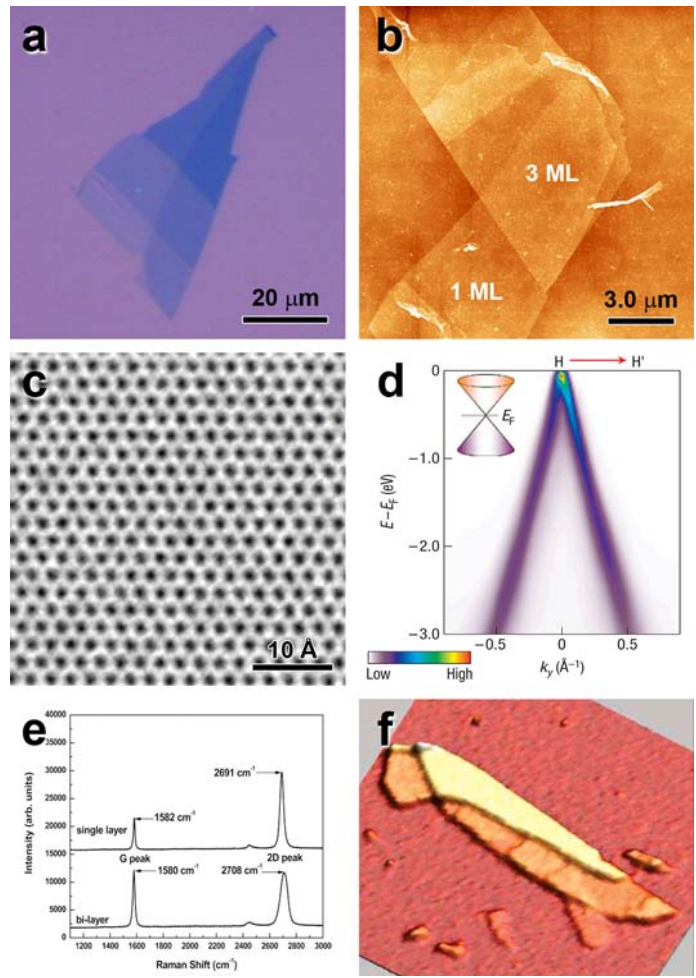

**Figure 3**

**Figure 3.** Major techniques for graphene characterization. (a) Optical microscopy, (b) Atomic Force Microscopy (AFM), (c) Transmission Electron Microscopy (TEM), (d) Angle-Resolved Photoemission Spectroscopy (ARPES), (e) Raman scattering and (r) Rayleigh scattering. Adapted with permission from (c) [23], copyright 2008 American Chemical Society; (d) [27], MacMillan Publishers Ltd: Nature Physics, copyright (2006); (e) [41], copyright 2007 American Chemical Society; (f) [42], copyright 2007 American Chemical Society;

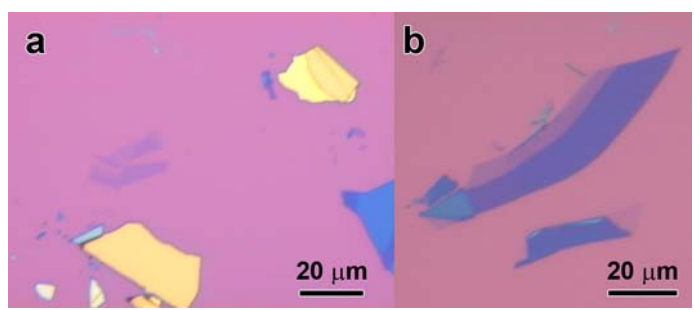

**Figure 4**

**Figure 4.** Micromechanically exfoliated graphene. Optical images of (a) thin graphite and (b) few layer graphene (FLG) and single layer graphene (lighter purple contrast) on a ~300 nm $SiO_2$ layer. Yellow-like color indicates thicker samples (~100s of nm) while bluish and lighter contrast indicates thinner samples.

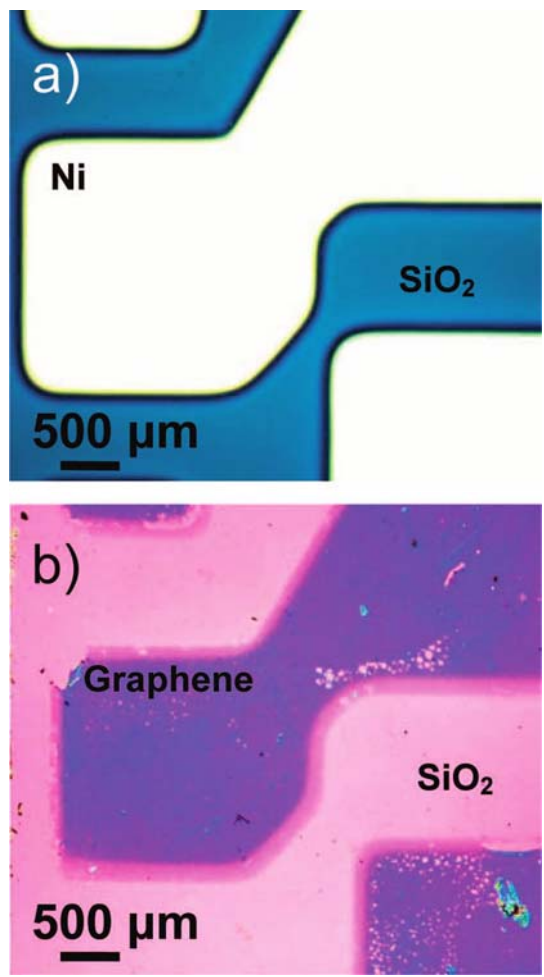

**Figure 5**

**Figure 5.** Nickel-grown graphene. (a) Optical image of a pre-patterned Ni film on SiO$_2$/Si. CVD graphene is grown on the surface of the Ni pattern. (b) Optical image of the grown graphene transferred intact from the Ni surface in (a) to another SiO$_2$/Si substrate. Reproduced with permission from [58]. Copyright 2009 American Chemical Society.

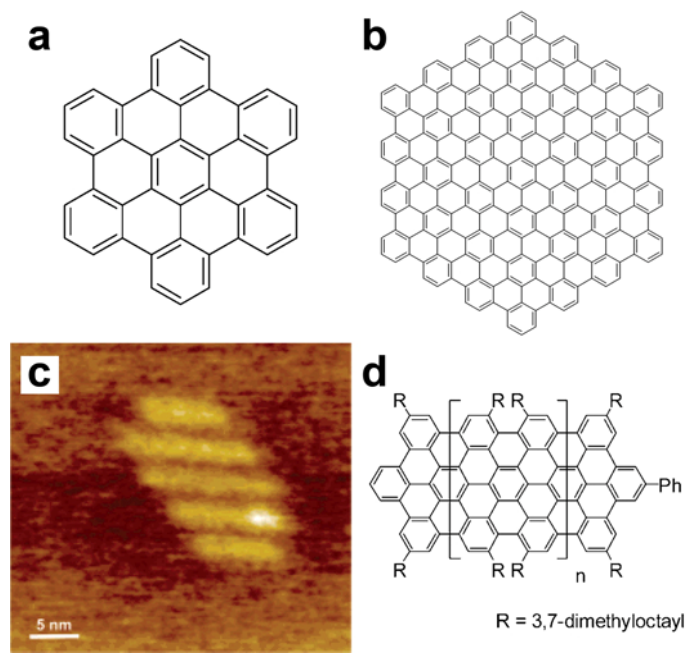

**Figure 6**

**Figure 6.** Molecular graphene. (a) Structure of the hexabenzocoronene (HBC). (b) Structure of the largest polycyclic aromatic synthesized to date, which contains 222 carbon atoms. (c) STM image at the solid/liquid interface on HOPG of self-assembled graphene nanoribbons shown in (d). Reproduced with permission from (a,b)[138], copyright 2007 American Chemical Society; (c,d) [137], copyright 2008 American Chemical Society.

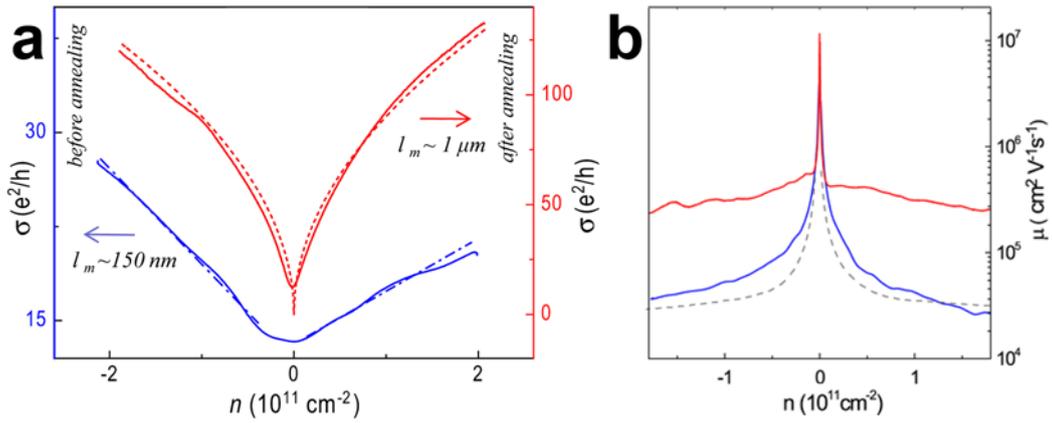

**Figure 7**

**Figure 7.** Carrier mobility in graphene. (a) Conductance of a suspended graphene sample (—, blue) before and (—, red) after annealing as a function of carrier density, n, at a temperature of 40 K. The black dotted line is the expected behavior for ballistic transport. (b) Mobility displayed as a function of n exceeds 200 000 cm$^2$/V.s at n = 2x10$^{11}$ cm$^{-2}$. Reproduced with permission from (a) [145], copyright 2008 APS and (b) [62], copyright (2008) from Elsevier.

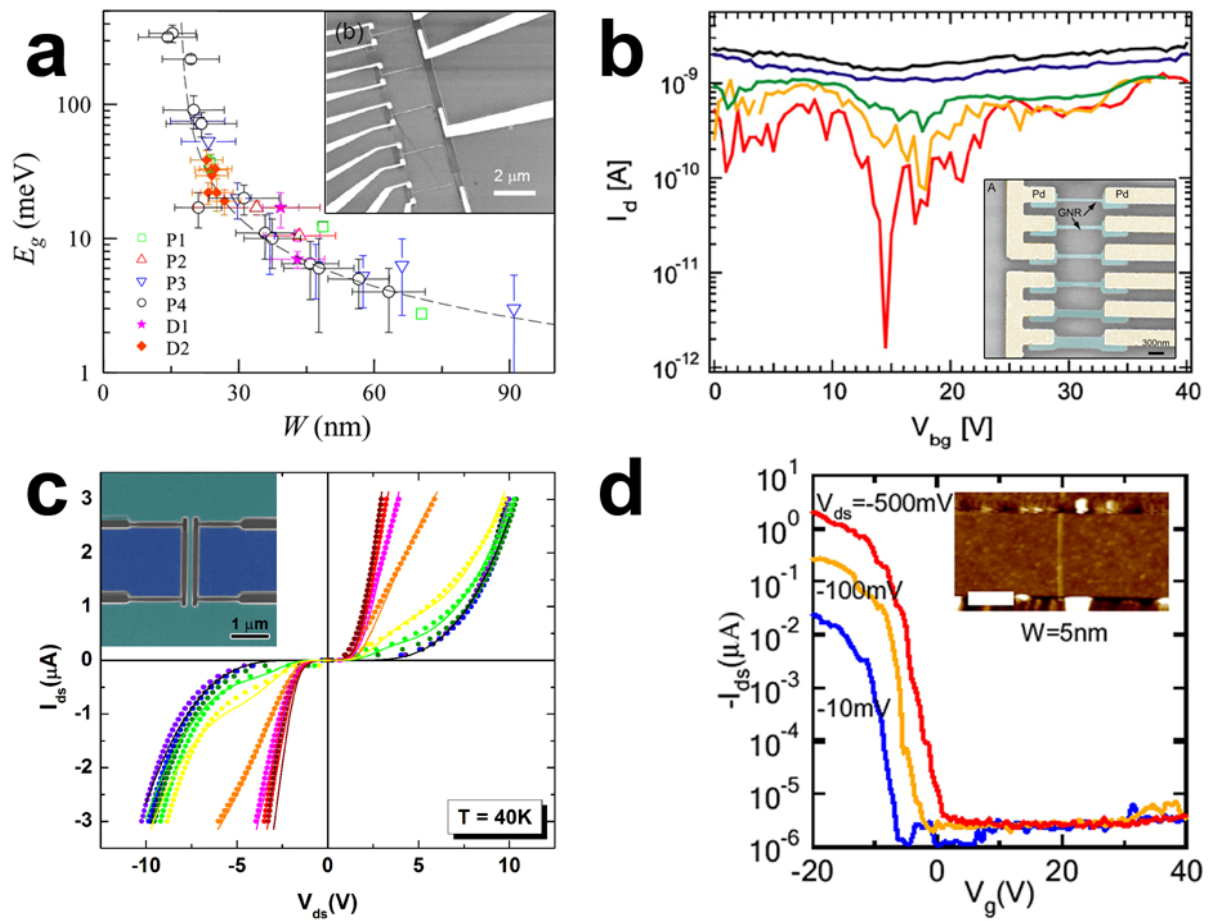

**Figure 8**

**Figure 8.** Graphene nanoribbon (GNR) field effect devices. (a) Apparent gap energy variation as a function of nominal GNR width (W < 100 nm) in backgated field effect devices obtained by lithography and ion etching. Inset: SEM image of a series of parallel GNR of different width. (b) Temperature dependence measurements of the transfer characteristics (source-drain current, $I_{ds}$, vs backgate voltage, $V_{gs}$) in a 20-nm wide GNR field effect device at source-drain voltage fixed at $V_{ds}$ = - 1mV. Temperatures are 400 K (black), 200 K (blue), 100 K (green), 50 K (orange), 4 K (red). The minimum current through the nanoribbon decreases by more than 1.5 orders of magnitude. Inset: False-colored SEM image of GNR devices fabricated by lithography and ion etching. The GNR widths from top to bottom are 20, 30, 40, 50, 10 and 200 nm. (c) Source-drain I-V characteristics at 40 K of a 50x2000 nm double side-gated nanoribbon with side gate voltage varying between 0 V (purple) and 30 V (dark red). Inset: False-colored SEM image of the few-layer graphene nanoribbon device with graphene source, drain and ribbon (cyan) as well as side gates (blue) etched by Focused Ion Beam. (d) Room-temperature transfer characteristics (current versus gate voltage $I_{ds}$-$V_{gs}$) for a 9 nm wide and 130 nm long bilayered GNR field effect device with Pd contacts and Si backgate. Inset: AFM image of the device. Scale bar is 100 nm. Reproduced with permission from (a) [126], copyright 2007 APS; (b) [197], copyright (2007) from Elsevier Science; (c) [60], Copyright Wiley-VCH Verlag GmbH & Co. KGaA;; (d) [120], copyright 2008 AAAS.

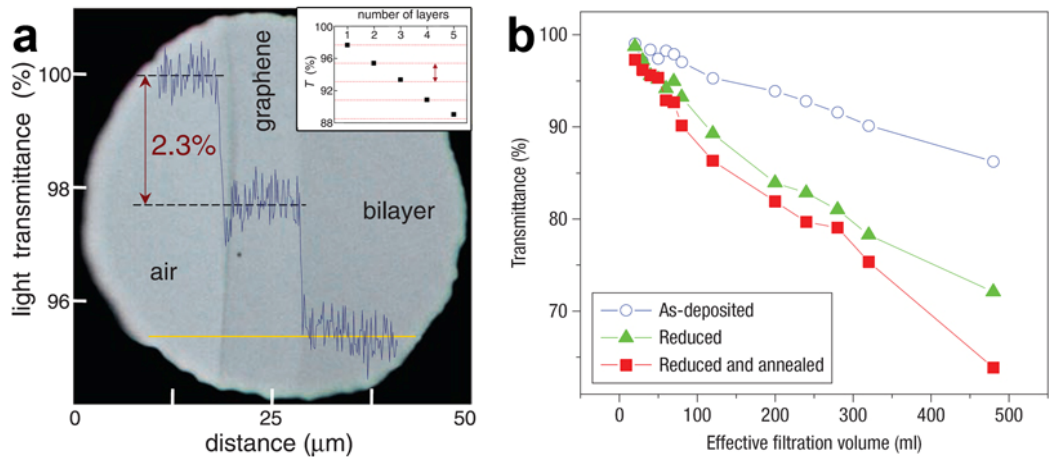

**Figure 9**

**Figure 9.** Optical transparency. (a) Optical micrograph of one- and twp-atom thick graphene crystals. The optical transmittance at 550 nm as a function of the lateral coordinate shows clear steps, the height of which is proportional to the hyperfine constant. In inset, the linear variation of the transparency as function of number of layers is observed up to six layers. (b) Transmittance at λ~550 nm as a function of the thickness of reduced GO thin films, assessed indirectly by the total volume of filtered suspension. Plots are shown for thin films with different reduction steps. From [201], Reprinted with permission from AAAS; (b) Reproduced with permission from [109], MacMillan Publishers Ltd: Nature Nanotechnology, copyright (2008)

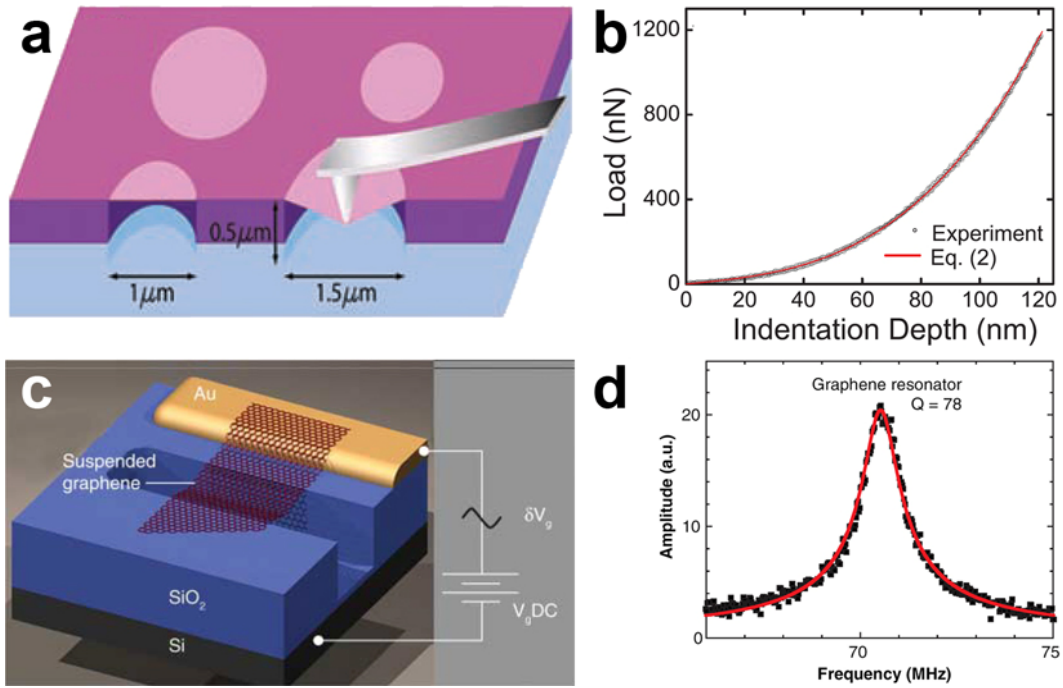

**Figure 10**

**Figure 10** Mechanical properties. (a) Schematic of AFM nanoindentation on suspended graphene. (b) Loading/unloading curve for device in (a) with modeling comparison (red line). (c) Schematic of a suspended graphene resonator. (d) Amplitude *vs.* frequency resonance curve taken with optical drive for the fundamental mode of a single-layer graphene resonator. (a) From [208]; (b) from [214], Reprinted with permission from AAAS.